\documentclass[a4paper,fleqn,usenatbib]{mnras}

\usepackage{newtxtext,newtxmath}
\usepackage{txfonts}

\usepackage[T1]{fontenc}
\usepackage{ae,aecompl}
\usepackage{pdflscape}

\usepackage{graphicx}	% Including figure files
\usepackage{amsmath}	% Advanced maths commands
\usepackage{amssymb}	% Extra maths symbols

   \title{VVV High proper motion stars I. The catalogue of bright $K_{\rm S}\le 13.5$ stars}

   \author[R. Kurtev et al.]{R. Kurtev$^{1,2}$\thanks{Email: radostin.kurtev@uv.cl}, 
M. Gromadzki$^{2,1}$, 
J. C. Beam\'in$^{2,1}$,
S. L. Folkes$^{1,8}$, 
K. Pena Ramirez$^{2,3}$,
\newauthor
V. D. Ivanov$^{4,5}$,
J. Borissova$^{1,2}$,
V. Villanueva$^{2,1}$,
D. Minniti$^{6,2}$,
R. Mendez$^{7,2}$,
P. W. Lucas$^{8}$,
\newauthor
L. C. Smith$^{8}$,
D. J. Pinfield$^{8}$,
M. A. Kuhn$^{2,1}$,
H. R. A. Jones$^{8}$,
A. Antonova$^{9}$,
and A. K. P. Yip$^{2,1}$\\
$^{1}$Instituto de F\'isica y Astronom\'ia, Universidad de Valpara\'iso, Av. Gran Breta\~na 1111, Playa Ancha, Casilla 5030, Valpara\'iso, Chile\\
$^{2}$Millennium Institute of Astrophysics, Santiago, Chile \\
$^{3}$Instituto de Astrof\'isica, Facultad de F\'sica, Pontificia Universidad Cat\'olica de Chile, Casilla 306, Santiago 22, Chile\\
$^{4}$European Southern Observatory, Ave. Alonso de Cordoba 3107, Casilla 19001, Santiago, Chile\\
$^{5}$European Southern Observatory,Karl-Schwarzschild-Str. 2, D-85748 Garching bei M\"unchen, Germany\\
$^{6}$Departamento de Ciencias F\'isicas, Universidad Andres Bello, Republica 220, Santiago, Chile\\
$^{7}$Universidad de Chile, Departamento de Astronom\'ia, Casilla 36-D, Santiago, Chile\\
$^{8}$Centre for Astrophysics Research, University of Hertfordshire, Hatfield AL10 9AB, UK\\
$^{9}$Department of Astronomy, Faculty of Physics, Sofia University, 5 James Bourchier Blvd., 1164 Sofia, Bulgaria
}

\date{}
\pubyear{2015}

\begin{document}
\label{firstpage}
\pagerange{\pageref{firstpage}--\pageref{lastpage}}
\maketitle

\begin{abstract}
Knowledge of the stellar content near the Sun is important for a broad range of topics ranging from the search for planets to the study of Milky Way structure. The most powerful method for identifying potentially nearby stars is proper motion (PM) surveys. All old optical surveys avoid, or are at least substantially incomplete, near the Galactic plane. The depth and breadth of the ``Vista Variables in V\'ia L\'actea'' (VVV) near-IR survey significantly improves this situation. Taking advantage of the VVV survey database, we have measured PMs in the densest regions of the MW bulge and southern plane in order to complete the census of nearby objects. We have developed a custom PM pipeline based on VVV catalogues from the Cambridge Astronomy Survey Unit (CASU), by comparing the first epoch of $JHK_{\rm S}$ with the multi-epoch $K_{\rm S}$-bands acquired later. Taking advantage of the large time baseline between the 2MASS and the VVV observations, we also obtained 2MASS-VVV PMs. We present a near-IR proper motion catalogue for the whole area of the VVV survey, which includes 3003 moving stellar sources. All of these have been visually inspected and are real PM objects. Our catalogue is in very good agreement with the proper motion data supplied in IR catalogues outside the densest zone of the MW. The majority of the PM objects in our catalogue are nearby M-dwarfs, as expected. This new database allow us to identify 57 common proper motion binary candidates, among which are two new systems within 30~pc of the Sun.
\end{abstract}


%%%%%%%%%%%%%%%%%%%%%%%%%%%%%%%%%%%%%%%%%%%%%%%%%%

%%%%%%%%%%%%%%%%% BODY OF PAPER %%%%%%%%%%%%%%%%%%

\section{Introduction}

A complete census of stars within the solar neighbourhood out to a specified distance will inform us about the stellar mass function, star formation, and the kinematics of the Galaxy and of young, nearby clusters and moving groups. The main difficulty in constructing a volume-limited sample is identification of nearby, low-mass objects because of their low luminosity. Also, accurate distance measurements for these stars are not easy to obtain.

The most powerful method for identifying potential nearby stars comes from PM surveys. PM surveys continually improve, as longer time baselines increase the accuracy of the measurement. A uniform census of nearby stars allows characterisation of the relative occurrence rates of different types of stars, and allows relationships between intrinsic properties of those stars, including absolute magnitude and colour, to be examined. 

The first attempts at large surveys for high proper-motion (HPM) stars began in the early 20th century with works by \citet{1915PASP...27..240V}, \citet{1919VeHei...7..195W} and \citet{1939AJ.....48..163R}. Later, additional surveys were completed: e.g., \citet{1971lpms.book.....G, 1978LowOB...8...89G}. The first all-sky search exploiting these initial photographic surveys for nearby stars was by Luyten, who published two PM catalogues: the Luyten Half-Second catalogue \citep[LHS:][]{1979lccs.book.....L}, and the New Luyten Two-Tenths catalogue \citep[NLTT:][]{1979nlcs.book.....L, 1979NLTT..C01....0L, 1980nltt.bookQ....L, 1980nltt.bookR....L}. 

After these early works many papers concerning proper motion studies were published. \cite{2005AJ....129.1483L} compiled a list of 61977 stars in the northern hemisphere with $\mu > 0.15$ arcsec yr$^{-1}$, identifying over 90\% of those stars down to a limiting magnitude of V$\approx$19.0, excluding the Galactic plane, and southern hemisphere (\citet{2008AJ....135.2177L}, $0.45 \le \mu \le 2.0\,\arcsec{\rm yr^{-1}}$). \citet{2005AJ....130.1247L} reported the discovery of 182 southern stars ($\delta < -30^\circ$) with proper motion $0.45 < \mu < 2.0\,\arcsec{\rm yr^{-1}}$. \cite{2005AJ....130.1658S} reported the discovery of 152 new high proper motion systems ($\mu\ge 0.4\,\arcsec {\rm yr^{-1}}$) in the southern sky ($\delta=-47^\circ$ to $00^\circ$) brighter than UKST plate $\rm R_{59F}=16.5$ via their SuperCOSMOS-RECONS (SCR) search.  \cite{2008AJ....135.2177L}, completing their SUPERBLINK proper motion survey in the southern celestial hemisphere, found 170 additional new stars with proper motions $0.45 < \mu < 2.0\,\arcsec{\rm yr^{-1}}$. This final part of their search covers 11,600 deg$^{-2}$ in the declination range $-30^\circ < \delta < 0^\circ$  and in low Galactic latitude areas south of $\delta$ = -30$^\circ$ which had not been covered in earlier data releases. Most of the new discoveries were found in densely populated fields along the Milky Way, toward the Galactic bulge/center. Their total list of high proper motion stars recovered by SUPERBLINK in the southern sky contains 2228 stars with proper motions  $0.45 < \mu < 2.0\,\arcsec{\rm yr^{-1}}$.

\cite{2011AJ....142...10B}, as a continuation of the SCR search in the southern sky, presented 2817 new southern proper motion systems with $0.18 < \mu < 0.40\,\arcsec{\rm yr^{-1}}$ and declinations between $-47^\circ$ and $00^\circ$. Subsequently, \cite{2011AJ....142...92B} published 1584 new southern proper motion systems with $\mu > 0.18\,\arcsec{\rm yr^{-1}}$ and $16.5 > R_{59F} \geq 18.0$. This search complemented the six previous SCR searches of the southern sky for stars within the PM motion range, but shallower than $R_{59F} = 16.5$.

\cite{2011AJ....142..138L} published an all-sky catalogue of M-dwarfs with apparent infrared (IR) magnitude $J<10$. The 8889 stars were selected from the on-going SUPERBLINK survey of stars with $\mu>40$~mas yr$^{-1}$, supplemented on the bright end with the Tycho-2 catalogue. Completeness tests suggest that this catalogue represents $\sim$75\% of the estimated $\sim$11900 M-dwarfs with $J<10$ expected for the entire sky. This catalogue is, however, significantly more complete for the northern sky ($\approx$90\%) than it is for the south ($\approx$60\%). \citet{2011AJ....141...97W} presented a spectroscopic catalogue of 70841 visually inspected M-dwarfs from the seventh release of the Sloan Digital Sky Survey. Recently, \citet{2013AN....334..176L}, again using the SUPERBLINK PM survey, reported a catalogue of $\sim$200,000 M-dwarfs in the northern sky. They presented a new census of $\sim$100,000 M-dwarfs located within 100 pc. The new census is 95\% complete to 50 pc, and $>$75\% complete to 100~pc. It was followed by the spectroscopic catalogue of the brightest ($J$$<$9) M dwarf candidates in the northern sky \citep{2013AJ....145..102L}.  \citet{2013MNRAS.435.2161F} used the Position and Proper Motion Extended-L (PPMXL) catalogue \citep{2010AJ....139.2440R} and applied optical and near-IR colour cuts together with a reduced proper motion (RPM) cut to find bright M-dwarfs for future exoplanet transit studies. Recently, \citet{2014MNRAS.437.3603S,2014MNRAS.443.2327S} presented two new infrared PM catalogues based on the UKIDSS Large Area Survey and Galactic Plane Survey.  

\citet{2014ApJ...781....4L} used multi-epoch astrometry from the Wide-field Infrared Survey Explorer (WISE) to identify 762 high proper motion objects, 761 of which  were detected also by the Two Micron All Sky Survey (2MASS). \citet{2014ApJ...783..122K}, using the AllWISE processing pipeline, have measured motions for all objects detected on WISE images taken between 2010 January and 2011 February. They found 22445 objects that have significant AllWISE motions, of which 3525 have motions that can be independently confirmed from earlier 2MASS images, yet lack any published motions in SIMBAD. 

M-dwarfs are the most abundant inhabitants of our Galaxy and also are probably the most common sites of planet formation \citep{2006ApJ...640L..63L}. They account for over 70\% of stellar systems in the solar neighbourhood \citep{1997AJ....114..388H}. In addition, the single star fraction -- a crucial statistic for giant planet formation \citep[e.g.][]{2012ApJ...745...19K} -- decreases from $\sim$60-70\% for M-dwarfs \citep{1992ApJ...396..178F, 2010A&A...520A..54B} to $\sim$54\% for solar-type stars \citep{1991A&A...248..485D, 2010ApJS..190....1R} to near 0\% for the most massive stars \citep{1999NewA....4..531P}, further separating M-dwarfs from AFGK stars as the most numerous potential planet hosts of all the stellar classes \citep{2006ApJ...640L..63L}. About 25\% of all Doppler-confirmed planets with $M \sin i < 30$~$M_\oplus$ are orbiting M-dwarfs. Large exoplanet surveys have now started to monitor sizable numbers of M-dwarfs, such as the M2K program which is targeting some 1600 M-dwarfs for radial velocity (RV) monitoring \citep{2010PASP..122..156A} and the MEarth project \citep{2014arXiv1409.0891I} which is designed to detect exoplanet transits in nearby late-type M-dwarfs. The principal methods of exoplanet detection, RV and transits, are both more sensitive to planets around stars of lower and substellar mass. The brightest M-dwarfs are the ideal (highest priority) targets for high precision RV searches, the latest M-dwarfs are the most suitable for transit surveys, and the youngest M-dwarfs are preferable targets for Adaptive Optics (AO) imaging. The above-mentioned factors make searches for new, nearby, and young low-mass stars and substellar objects highly valuable.

All old optical surveys avoid or are at least substantially incomplete near to the Galactic plane. In the Subsection\,\ref{sec:catalog} we present a detailed comparison with previous PM studies covering the VVV area. There are two modern surveys that make an exception: The INT Photometric H$_\alpha$ Survey of the Northern Galactic Plane (IPHAS) \citep{2005MNRAS.362..753D} and The VST Photometric H$_\alpha$ Survey of the Southern Galactic Plane and Bulge (VPHAS+) \citep{2014MNRAS.440.2036D}. Nevertheless, this region is still referred to as the ``zone of avoidance'' as it contains the highest stellar densities down to faint limiting magnitudes in addition to regions with dark molecular clouds, nebulosity, and current star formation, which produces substantial confusion. Nevertheless, the Galactic plane and the Galactic bulge offer considerable latent potential for new discoveries of nearby low-mass stars and ultra-cool dwarfs (UCDs) from deeper searches. Such discoveries may contain young, unusual, and nearby/bright examples of these objects, and will also complement those made at higher Galactic latitudes \citep{2012MNRAS.427.3280F}. Good recent examples of nearby ($<$10 pc) interesting discoveries at low Galactic latitude are those of UGPS~J0722-05 \citep[T9:][]{2010MNRAS.408L..56L}, DENIS~J081730.0-615520  \citep[T6.5:][]{2010ApJ...718L..38A}, and the amazing discoveries of the nearest brown dwarfs  WISE~J104915.57-531906.1AB \citep{2013ApJ...767L...1L} and WISE~J085510.83-071442.5 \citep{2014ApJ...786L..18L} at $\sim$2~pc from the Sun. These discoveries near to the Galactic plane highlight the important serendipitous nature in which new IR surveys like VVV \citep{2010NewA...15..433M} can improve on the incomplete Solar neighbourhood census of low-mass stellar and sub-stellar systems. 

A positive aspect of the high stellar densities encountered in the Galactic plane and bulge is the plethora of bright reference stars for good AO tip-tilt low-order correction. This will facilitate high-Strehl imaging measurements to identify very low-mass brown dwarf/planetary-mass, companions for studying multiplicity, and also measuring dynamical masses. Moreover, high-Strehl AO studies of newly identified binary moving group members could also provide good age and composition constrains, as well as dynamical masses, enabling direct feed-back to evolutionary models. The crowded fields are also perfect for time series and astrometric studies because they provide many suitable reference stars.

The VVV near-IR ($ZY JHK_{\rm S}$) survey \citep{2010NewA...15..433M, 2012A&A...544A.147S} covers 562 deg$^2$ of the Galactic bulge and the Southern Galactic disk and provides accurate photometry, and multi-epoch $K_{\rm S}$-band imaging, enabling us to discover a meaningful sample of new nearby cool and ultra-cool dwarfs (UCDs: spectral types $>$M6) with a higher completeness than has previously been achieved in the low Southern Galactic latitudes (e.g., from 2MASS and Deep Near Infrared Survey of the Southern Sky (DENIS)), from a PM search. The first study of the PM objects using VVV data was made by \citet{2013AA...560A..21I}. The common proper motion method was used and seven new co-moving companions around known HPM stars were discovered. \citet{2013A&A...557L...8B} reported the discovery of the first VVV brown dwarf -- VVV\,BD001 -- a new member of the 20 pc sample with well defined proper motion, distance, and luminosity. For our initial search we limited our sample selection to the brightest ($K_{\rm S}<13.5$ magnitude) objects only. A more complete and deeper catalogue will be published later \citep[Smith et al., in prep]{smith2015a}.

Here we present the first VVV HPM catalogue limited to $K_{\rm S}\le13.5$ using the VVV databases.

   \section{Proper Motion Candidate Selection}
   \label{sec:methods}

Initially, we identify suitable VVV FITS catalogues to download and analyse from the CASU, by selecting the near-IR multi-colour photometrically deeper dataset (first epoch: $JHK_{\rm S}$)~and the separate multi-epoch $K_{\rm S}$-bands (hereafter, Kv epochs) of interest for an individual tile. We choose Kv epoch frames that have similar $5\sigma$~photometric detection limits as well as having similar r.m.s.\ residuals to the 2MASS-based astrometric-fit uncertainties (CASU calculated parameter), to aid in the proper identification of the same objects over all the epochs in order to reduce the incidence of contamination and false-positive detections.

A positional cross-match is then made on the celestial coordinates on the first epoch $JHK_{\rm S}$~FITS catalogue data using {\sc stilts}-v2.3 \citep[http://www.starlink.ac.uk/stilts/]{2006ASPC..351..666T} with a 0.3~arcsec matching radius. We also add a requirement that all valid sources for inclusion must have detections in all three $JHK_{\rm S}$~catalogues. These multi-colour datasets are contemporaneous in epoch for a given tile, and allow us to obtain near-IR colours for any PM candidates we identify.  The cross-matched detections are filtered during the conversion by the use of the CASU morphological classification flag: sources are only selected having flag attribute values of -9, -2 and -1 (saturated, borderline stellar, and stellar, respectively).

\subsection{Initial Selection}

We read temporal keyword metadata from the catalogue FITS headers and calculate the search radius limit in arcseconds to be applied in the cross-match between detections in the catalogues with differing epochs. The cross-match is based on a 5~arcsec~yr$^{-1}$~maximum PM limit that we apply, as well as a constraint ($\Delta K_{\rm S}\leq0.28$~mag) placed on the difference between the two $K_{\rm S}$-band magnitudes of cross-matched detections. This photometric constraint is based on the relaxed (0.2~mag) absolute $3\sigma$~photometric uncertainty given by CASU of 0.15 magnitudes in the $K_{\rm S}\sim 14$ for the VVV survey (Saito et al.\ 2012), and calculated as the $\sqrt{0.2^2 + 0.2^2}$ for two separate detections being compared. This is applied to help match the correct source in fields of higher source density. 

After the cross-match has been made to identify possible motions of up to 5~arcsec~yr$^{-1}$, a constraint is then placed on a minimum allowable separation for valid candidate inclusion, which is based on the r.m.s.\ astrometric uncertainties of the equatorial coordinates in the particular catalogues (CASU derived values). This constraint requires that, for motion to be declared valid between the two catalogues, the measured separation must have a value $\ge\sqrt{(RMS_1)^2 + (RMS_2)^2}$, where the RMS values are in arcseconds. The combined minimum separation constraint used here implies that a lower limit to the PM sensitivity exists that is also dependent on the temporal separation between epochs used, and therefore, will vary slightly between tiles.

During later testing of our method it was found that rather than using a $3\sigma$~astrometric uncertainty constraint for both the initial and refined stages of the analysis, relaxing this requirement for the initial candidate selection to a minimum of $1\sigma$~(the r.m.s.\ value; typically about 0.07 arcsec) allowed us to reliably detect PMs down to our lower sensitivity limits. Therefore, we adopted those values throughout our search.

\subsection{Second, Refined Selection}

Any cross-matched sources found to have photometric detections at individual epochs that are fainter than the brightest $5\sigma$~photometric limit of all the measured Kv epochs per tile, are removed as PM candidates. This is to reduce the instances of object miss-matches and avoid possible positional inaccuracies associated with fainter detections. 

For each PM candidate identified from the initial selection, a test is then conducted to establish if both the PM and position angle measurements (separation vectors) between all the successive cross-matched epoch pairs are consistent with each other. The test is based on the assumption that any detectable motion should be linear, and is conducted in reverse temporal order of epoch with each separation vector measurement for each candidate required to be consistent to pass the test.

To check that successive pairs of separation vectors are consistent, the separation vector from the two latest measured Kv epoch pairs (say, epochs 3 and 2) is used to predict the position of a given PM candidate at epoch 1 (assuming three epochs are available). If the measured position and predicted position are within a calculated maximum test radius (depending on the positional errors and epoch difference) then the PM candidate is retained. If three separate epochs are used (minimum required) then the test can only be conducted once, however, for each successive epoch added another test is also made, thereby increasing the effectiveness of the test in rejecting contamination.

As mentioned above we relaxed the astrometric constraint in the initial selection, while using a $3\sigma$~test radius when comparing the separation of the predicted positions with that of the measured positions, to test for consistent separation vectors. Due to the use of these differing astrometric constraints, there exists the possibility that for candidates passing with the smallest accepted separations (smallest PMs), their position angles and motion could be unrepresentative of a true underlying PM (i.e., spurious detections). To alleviate this possibility, we included an additional test in the refined analysis to check for realistic and consistent measurements in position angle over the separate epochs.

This position angle test uses a geometrical argument that for two successive pairs of separation vectors (with the minimum possible separation) the corresponding positions measured at three separate epochs is likely to take the form of a triangle, but is unlikely to result in a triangle with an internal angle to the apex of $<60^\circ$, i.e. with an angle smaller than that of an equilateral triangle. The assumption here being that the apparent change in positions is due to true underlying linear PM. If the two sides of a triangle formed by a pair of separation vectors
must form an angle >60 degrees, then the allowable difference in
position angle would have to be $\mbox{PA}\leq120^\circ$. This angle is the constraint we impose on successive position angle measurements for inclusion as a valid PM candidate.
It was often noticed during the testing of our method that close blended sources in crowded areas, often associated with variability or regions of poorer astrometric fit, which were coincident over a small number of available epochs, would pass as PM candidates. To facilitate the removal of these false-positives, we introduced a final test on the candidates during the refinement stage of the analysis. This took the form of a cross-match of the candidate positions from the latest epoch, with the first epoch catalogue (the original multi-colour dataset). If the candidate is matched within a test radius in the first epoch, and it has a similar apparent magnitude within $\Delta K_{\rm S}\leq0.28$~(as used in the initial selection above), then it is rejected. The test radius used to determine valid PM in this case is defined by: the difference between positions must be greater than the PM lower sensitivity limit (in arcseconds) multiplied by the temporal difference between the epochs.

For the remaining candidates that pass all test criteria we use the mean values of their PM and position angle measurements obtained over all their measured epochs, and these were used as the initial values for the search. The
final values of the VVV PMs in the catalogue, however, were obtained
using the equatorial coordinates in the first and the last available VVV epoch.

Finally, to reduce the number of PM candidates from all the VVV tiles to a manageable number for visual checking and cross-matching with other survey data such as 2MASS, we introduced a photometric cut ($K_{\rm S}\leq13.5$~mag) to select only the brightest candidates.

We summarise the main points of our VVV PM candidate selection method as follows:

\begin{enumerate}
\item Select, download, and decompress all FITS catalogues from CASU per tile.
\item Cross-match the first epoch multi-colour $JHK_{\rm S}$ FITS catalogue dataset.
\item Select sources with ``CASU classification flag'': values of -9, -2 and -1 only.
\item Convert FITS results file to ASCII for analysis.
\item Repeat a cross-match and initial analysis on a per-epoch basis per tile, to identify PM candidates.
\item Make a refined analysis over all epochs: perform a test for consistent motion and position angle. 
\item Apply a test to facilitate the removal of false-positive detections.
\item Make a selection of the brightest PM candidates of $K_{\rm S}\leq13.5$~magnitude for visual checking and follow-up.
\end{enumerate}

\subsection{Final Visual Inspection}

As it was already mentioned above, it is very difficult to operate a PM search in such crowded fields of stars, with too many contaminants provoking the detection of numerous false HPM objects. The main source of errors are the bright heavily saturated stars which produce numerous peaks and spikes in a surrounding area. In some cases they are too numerous and easily some of them could fulfil the above-mentioned search and refine criteria. The other significant source of errors comes from some very close stellar doublets with similar magnitudes. All these factors made a final visual check mandatory. We visually examined around 20000 probable HPM objects. The first step was to blink on the DS9 the first and last available VVV epoch  ($1 \times 1$~armin) images around each candidate available in the Vista Science Archive (VSA) \footnote{http://horus.roe.ac.uk/vsa/index.html} and to check for apparent motion. The time span between these epochs in most cases was about four years. Then we overplotted the position of the 2MASS stars in the area and checked again for the apparent motion. The limiting magnitude of $K_{\rm S}=13.5$\,mag for our sample is relatively bright and the majority of the stars have a 2MASS counterpart. We have provided a visual check for all candidates and the final list contains only visually confirmed PM objects.

\section{Results}
\label{sec:results}

\subsection{The PM Catalogue}
\label{sec:catalog}

We find 3003 stars with PM $>$30 mas\,yr$^{-1}$ in the entire VVV area (bulge and disk). The catalog of these objects is provided in Table\ref{tab1}, which is published in full in electronic format only. In the table, stars are ordered by right ascension. The columns of this table are:
\begin{enumerate}
\item Consecutive number
\item Right ascension (VVV first epoch) in degrees
\item Declination (VVV first epoch) in degrees
\item Epoch (VVV first epoch), JD
\item Galactic longitude in degrees
\item Galactic latitude in degrees
\item VVV $Z$ magnitude
\item Error of the VVV $Z$ magnitude
\item VVV $Y$ magnitude
\item Error of the VVV $Y$ magnitude
\item VVV $J$ magnitude
\item Error of the VVV $J$ magnitude
\item VVV $H$ magnitude
\item Error of the VVV $H$ magnitude
\item VVV $K_{\rm S}$ magnitude
\item Error of the VVV $K_{\rm S}$ magnitude
\item Right ascension (2MASS) in degrees
\item Declination (2MASS) in degrees
\item Epoch (2MASS), JD
\item 2MASS $J$ magnitude
\item Error of the 2MASS $J$ magnitude
\item 2MASS $H$ magnitude
\item Error of the $H$ 2MASS magnitude
\item 2MASS $K_{\rm S}$ magnitude
\item Error of the 2MASS $K_{\rm S}$ magnitude
\item $\mu_\alpha$(VVV) in arsec\,yr$^{-1}$
\item $\mu_\delta$(VVV) in arsec\,yr$^{-1}$
\item $\mu$(VVV) in arsec\,yr$^{-1}$
\item PA(VVV)  in degrees
\item $\mu_\alpha$(2MASS) in arsec\,yr$^{-1}$
\item $\mu_\delta$(2MASS)  in arsec\,yr$^{-1}$
\item $\mu$(2MASS) in arsec\,yr$^{-1}$
\item PA(2MASS) in degrees
\item Comment 
\end{enumerate}

Columns (ii, iii) give the first-epoch VVV coordinates and columns (xvii, xviii) give the 2MASS coordinates. Columns (iv) and (xix) give the corresponding epochs. Right ascension ($\alpha$) and declination ($\delta$) are listed in the ICRS system. The current positions must be extrapolated using the tabulated PMs. This table gives the VVV $ZYJHK_{\rm S}$ magnitudes (VISTA system) as well as the 2MASS magnitudes of the objects. As discussed, the brighter stars (any VVV ${\rm mag} < 12.0$) are in the zone of saturation or non-linearity and their astrometry should be used with
caution. It is not possible to state the exact limit for saturation or non-linearity because it depends on the observational conditions during the observation of the corresponding VISTA paw print, the filter, the chip of the detector, etc. For this reason we decided to also give the 2MASS magnitudes of stars. On the other hand, for some of the fainter stars, even though our limit of $K_{\rm S}=13.5$ is significantly brighter than the 2MASS
magnitude limit of 16.0, the magnitudes are missing in the 2MASS catalogue. These are very few cases and are usually stars in crowded regions or near a bright star. In other cases a near neighbour is unresolved in the 2MASS images yielding incorrect 2MASS photometry. In these cases VVV magnitudes are repeated in the columns labeled ``2MASS'' rather than 2MASS photometry, and this is noted in the column ``Comment'' with {\sl VVV mag}. For these stars, the proper motion is obtained using only the VVV data.

\begin{landscape}
\begin{table}
\caption{The VVV HPM Catalogue. The full catalogue is available in electronic format.}\label{tab1}
\centering
\tabcolsep=1.5pt
\begin{tabular}{r|r|r|r|r|r|r|r|r|r|r|r|r|r|r|r|r|r|l}
\hline
    &&&&&&&&&&&&&&&&&\\[-7pt]
  \multicolumn{1}{c|}{No} &
  \multicolumn{1}{c|}{$\alpha_1$(VVV)} &
  \multicolumn{1}{c|}{$\delta_1$(VVV)} &
  \multicolumn{1}{c|}{JD$_1$(VVV)} &
  \multicolumn{1}{c|}{$l$} &
  \multicolumn{1}{c|}{$b$} &
  \multicolumn{1}{c|}{$Z$(VVV)} &
  \multicolumn{1}{c|}{Error $Z$} &
  \multicolumn{1}{c|}{$Y$(VVV)} &
  \multicolumn{1}{c|}{Error $Y$} &
  \multicolumn{1}{c|}{$J$(VVV)} &
  \multicolumn{1}{c|}{Error $J$} &
  \multicolumn{1}{c|}{$H$(VVV)} &
  \multicolumn{1}{c|}{Error $H$} &
  \multicolumn{1}{c|}{$K$(VVV)} &
  \multicolumn{1}{c|}{Error $K$} & 
  \multicolumn{1}{c|}{} &
  \multicolumn{1}{c|}{} &
  \multicolumn{1}{c}{} \\
  \multicolumn{1}{c|}{} &
  \multicolumn{1}{c|}{$\alpha$(2MASS)} &
  \multicolumn{1}{c|}{$\delta$(2MASS)} &
  \multicolumn{1}{c|}{JD(2MASS)} &
  \multicolumn{1}{c|}{$J$(2MASS)} &
  \multicolumn{1}{c|}{Error $J$} &
  \multicolumn{1}{c|}{$H$(2MASS)} &
  \multicolumn{1}{c|}{Error $H$} &
  \multicolumn{1}{c|}{$K$(2MASS)} &
  \multicolumn{1}{c|}{Error $K$} &  
  \multicolumn{1}{c|}{$\mu_\alpha$(VVV)} &
  \multicolumn{1}{c|}{$\mu_\delta$(VVV)} &
  \multicolumn{1}{c|}{$\mu$(VVV)} &
  \multicolumn{1}{c|}{PA(VVV)} &
  \multicolumn{1}{c|}{$\mu_\alpha$(2M)} &
  \multicolumn{1}{c|}{$\mu_\delta$(2M)} &
  \multicolumn{1}{c|}{$\mu$(2M)} &
  \multicolumn{1}{c|}{PA(2M)} &
  \multicolumn{1}{c}{Comment} \\
      &&&&&&&&&&&&&&&&&\\[-7pt]
\hline
  &&&&&&&&&&&&&&&&&\\[-7pt]
  1 & 174.141571 & -63.643942 & 2455226.723 & 294.736426 & -1.969864 & 12.838 & 0.001 & 12.468 & 0.001 & 11.905 & 0.001 & 11.330 & 0.001 & 11.019 & 0.001 & & & \\ & 174.142367 & -63.644070 & 2451584.743 & 11.997 & 0.027 & 11.311 & 0.026 & 11.111 & 0.023 & -0.128 & 0.042 & 0.135 & -71.8 & -0.128 & 0.045 & 0.136 & -70.6 & \\
  &&&&&&&&&&&&&&&&&\\[-7pt]
  2 & 174.246366 & -63.495102 & 2455269.617 & 294.738748 & -1.81391 & 13.856 & 0.001 & 13.350 & 0.001 & 12.768 & 0.001 & 12.254 & 0.001 & 11.989 & 0.002 & & & \\ & 174.247176 & -63.494961 & 2451584.743 & 12.825 & 0.023 & 12.222 & 0.029 & 11.906 & 0.031 & -0.120 & -0.036 & 0.125 & -106.7 & -0.127 & -0.047 & 0.135 & -110.3 & Comp A\\
    &&&&&&&&&&&&&&&&&\\[-7pt]
  3 & 174.248713 & -63.482605 & 2455226.723 & 294.736168 & -1.801644 & 14.360 & 0.002 & 13.831 & 0.002 & 13.229 & 0.002 & 12.706 & 0.001 & 12.426 & 0.002 & & & \\ & 174.249419 & -63.482475 & 2451584.743 & 13.248 & 0.050 & 12.655 & 0.042 & 12.358 & 0.031 & -0.118 & -0.050 & 0.128 & -113.0 & -0.115 & -0.048 & 0.125 & -112.7 & Comp B\\
    &&&&&&&&&&&&&&&&&\\[-7pt]
  4 & 174.280454 & -63.435237 & 2455226.723 & 294.736343 & -1.752237 & 14.193 & 0.002 & 13.704 & 0.002 & 13.151 & 0.001 & 11.944 & 0.001 & 12.486 & 0.002 & & & \\ & 174.281079 & -63.435143 & 2451584.743 & 13.164 & 0.032 & 12.565 & 0.045 & 12.138 & 0.037 & -0.093 & -0.024 & 0.096 & -104.5 & -0.098 & -0.031 & 0.103 & -107.6 & \\
    &&&&&&&&&&&&&&&&&\\[-7pt]
  5 & 174.369840 & -63.448234 & 2455269.617 & 294.778371 & -1.753348 & 12.540 & 0.001 & 12.330 & 0.001 & 11.920 & 0.001 & 11.701 & 0.001 & 11.635 & 0.001 & & & \\ & 174.370223 & -63.448421 & 2451584.749 & 11.959 & 0.026 & 11.644 & 0.023 & 11.520 & 0.021 & -0.049 & 0.065 & 0.081 & -37.0 & -0.058 & 0.066 & 0.088 & -41.3 & \\

  \multicolumn{19}{l}{... And 2998 more PM objects ...} \\
  \hline
\end{tabular}
\end{table}
\end{landscape}

\begin{figure}
\includegraphics[width=\columnwidth]{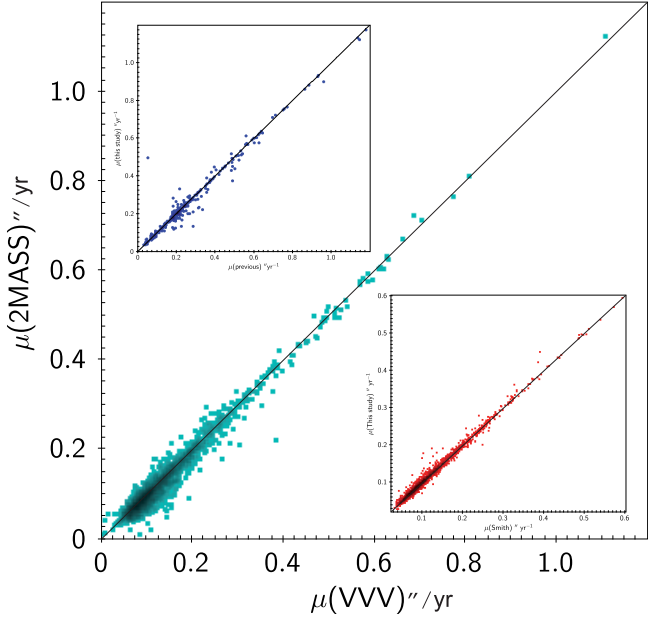}%
\caption{PM obtained from the (VVV1, VVV2) vs.\ (2MASS, VVV2). The lower inset (red points) gives  the comparison between the PMs obtained by this work versus those obtained by \citet{smith2015a}. The upper inset shows the comparison between the PMs for the previously known HPM objects in the VVV area obtained by this work and taken from the literature (SIMBAD). Both cases show that measurements using the different methods are consistent.}
\label{fig1}
\end{figure}

The final values of the PMs in the catalogue are obtained in two ways: a)~using the equatorial coordinates in the first VVV epoch (February/March 2010) and the last available VVV epoch (October 2014) with a typical separation in time of about four years, and b)~comparing the equatorial coordinates of the last available VVV epoch and 2MASS. These cases are referred to as (VVV1, VVV2) and (2MASS, VVV2), respectively. The majority of the targets have reliable 2MASS magnitudes and positions (except for some cases of near neighbours and some cases with missing data). The time separation between the epochs in this second case is always more than 10 years, reaching in some cases 15 years. The precision of the PM and the positional angle is similar using both procedures. In the first case, the precision of the VVV coordinates is much better, because of the survey's observational constraints and the smaller VISTA pixel scale, but the time difference is less. In the second, the lower precision of the 2MASS coordinates is compensated by 3 to 5 times longer delay between the epochs. The comparison of the PMs is given in Fig.\ref{fig1}. The coincidence of the results is very good. The mean dispersion of the PMs obtained in both ways is $\sigma$$\approx$$16$\,mas\,yr$^{-1}$, varying between 13\,mas\,yr$^{-1}$ for the $PM>300$\,mas\,yr$^{-1}$ and 23\,mas\,yr$^{-1}$ for PM<100\,mas\,yr$^{-1}$. 

We compared the PMs of the objects in our list with the PMs of the previously known HPM objects in the VVV area. First, we searched the SIMBAD database for the PM objects and cross-matched the obtained list with our catalogue. Because many sources are not listed in SIMBAD we also cross-matched our list with previous PM catalogues covering the VVV area. The details are given in Table\,\ref{tab2}. In the mean catalogue (Table\,\ref{tab1}), in the column ``Comment'', for each of these stars the corresponding designation and the catalogue reference are given. The result is presented in Fig.\ref{fig1}\,(upper inset). The dispersion along the line is 33 mas yr$^{-1}$. We also compared our results with those generated in the PhD thesis of Leigh Smith \citep{smith2015a} (see Fig.\ref{fig1}, lower inset). In this case the dispersion is only 7 mas yr$^{-1}$ which is an expected result because the study is based on the VVV data too. The detailed analysis and comparison will be published in the upcoming paper (Smith et al., in preparation).

\begin{landscape}
\begin{table}
\caption{Comparison with other catalogues covering the VVV area. The total number of the previously known PM stars is less than a sum of the stars in each catalogue because some of the stars are presented in more than one catalogue.}\label{tab2}
\centering
\begin{tabular}{r|l|l}
\hline
    &&\\[-7pt]
  \multicolumn{1}{c|}{No stars} & \multicolumn{1}{c|}{Catalogue/source} & \multicolumn{1}{c}{Reference} \\
    &&\\[-7pt]
\hline
 3 & First supplement to the NLTT catalogue. & \cite{1980PMMin..55....1L} \\
 5 & The TYCHO Reference Catalogue & \cite{1998AA...335L..65H} \\
 76 & The Tycho-2 catalogue of the 2.5 million brightest stars & \cite{2000AA...355L..27H} \\
 2& The Washington double star catalog & \cite{2001AJ....122.3466M} \\
 13 & Astrometry with the MACHO Data Archive. I. High Proper Motion Stars toward the Galactic Bulge and Magellanic Clouds & \cite{2001ApJ...562..337A} \\
 2 & New High Proper Motion Stars from the Digitized Sky Survey. I. Northern Stars with 0.5" yr$^{-1}<\mu<2.0"$ yr$^{-1}$ at Low Galactic Latitudes & \cite{2002AJ....124.1190L} \\
 5 & Revised Coordinates and Proper Motions of the Stars in the Luyten Half-Second Catalog & \cite{2002ApJS..141..187B} \\
 32 & Improved Astrometry and Photometry for the Luyten Catalog. II. Faint Stars and the Revised Catalog
 & \cite{2003ApJ...582.1011S}\\
 2 & VizieR Online Data Catalog: The Second U.S. Naval Observatory CCD Astrograph Catalog (UCAC2) & \cite{2003yCat.1289....0Z} \\
 12 & The Optical Gravitational Lensing Experiment: catalogue of stellar proper motions in the OGLE-II Galactic bulge fields & \cite{2004MNRAS.348.1439S}\\
 2 & New High Proper Motion Stars from the Digitized Sky Survey. III. Stars with Proper Motions $0.45" < \mu < 2.0"$ yr$^{-1}$ South of Declination $-30^\circ$ & \cite{2005AJ....130.1247L}\\
 1 & The Solar Neighborhood. XV. Discovery of New High Proper Motion Stars with $\mu \ge 0.4"$ yr$^{-1}$ between Declinations $-47^\circ$ and $00^\circ$ & \cite{2005AJ....130.1658S} \\
 33 & Validation of the new Hipparcos reduction & \cite{2007AA...474..653V} \\
 3 & The Solar Neighborhood. XVIII. Discovery of New Proper-Motion Stars with $0.40"\,{\rm yr}{-1} > \mu \ge 0.18"$ yr$^{-1}$ between Declinations $-90^\circ$ and $-47^\circ$ & \cite{2007AJ....133.2898F}\\
 21 & New High Proper Motion Stars from the Digitized Sky Survey. Iv. Completion of the Southern Survey and 170 Additional Stars with $\mu > 0.45''$ yr$^{-1}$ & \cite{2008AJ....135.2177L} \\
 41 & UCAC3 Proper Motion Survey. I. Discovery of New Proper Motion Stars in UCAC3 with $0.40"\,{\rm yr}{-1} > \mu \ge 0.18"$ yr$^{-1}$ between Declinations $-90^\circ$ and $-47^\circ$ & \cite{2010AJ....140..844F} \\
 2 & New Wide Common Proper Motion Binaries & \cite{2010JDSO....6...30B} \\ 
 24 & An All-sky Catalog of Bright M Dwarfs & \cite{2011AJ....142..138L} \\
 35 & The Solar Neighborhood. XXV. Discovery of New Proper-Motion Stars with $0.40"\,{\rm yr}{-1} > \mu \ge 0.18"$ yr$^{-1}$ between Declinations $-47^\circ$ and $00^\circ$ & \cite{2011AJ....142...10B} \\
 17 & The Solar Neighborhood. XXVII. Discovery of New Proper Motion Stars with $\mu \ge 0.18"$ yr$^{-1}$  in the Southern Sky with $\rm 16.5 < R_{59F} \le 18.0$ & \cite{2011AJ....142...92B} \\
14 & UCAC3 Proper Motion Survey. II. Discovery of New Proper Motion Stars in UCAC3 with $0.40"\,{\rm yr}{-1} > \mu \ge 0.18"$ yr$^{-1}$ between Declinations $-47^\circ$ and $00^\circ$ & \cite{2012ApJ...745..118F} \\ 
2 & VizieR Online Data Catalog: UCAC4 Catalogue (Zacharias+, 2012) & \cite{2012yCat.1322....0Z} \\
4 & Discovery of new companions to high proper motion stars from the VVV Survey & \cite{2013AA...560A..21I} \\
5 & A catalogue of bright (K < 9) M dwarfs & \cite{2013MNRAS.435.2161F} \\
1 & Nearby M, L, and T Dwarfs Discovered by the Wide-field Infrared Survey Explorer (WISE) & \cite{2013PASP..125..809T} \\
12 & A Search for a Distant Companion to the Sun with the Wide-field Infrared Survey Explorer
 & \cite{2014ApJ...781....4L} \\
 1 & Discovery of a brown dwarf companion to the A3V star $\beta$ Circini & \cite{2015MNRAS.454.4476S} \\
 && \\
\multicolumn{3}{l}{In total, 348 known PM object from all 3003 stars in our list ($\sim$11.5\%). } \\
 
\hline
\end{tabular}
\end{table}
\end{landscape}

\begin{figure}
\includegraphics[width=\columnwidth]{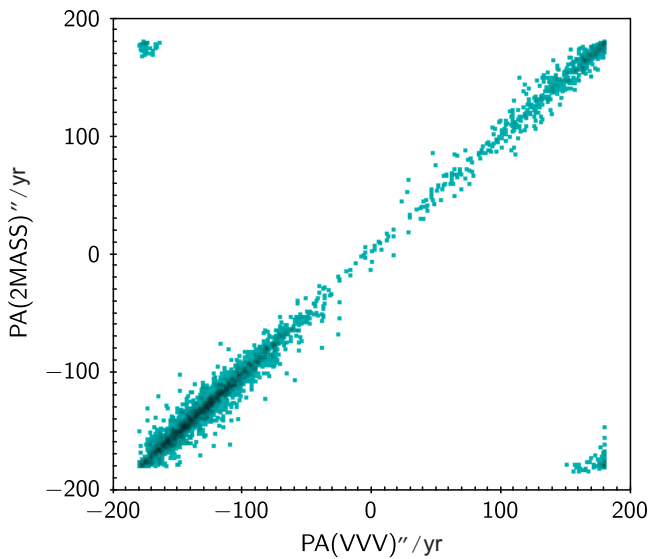}%
\caption{Positional angle (PA) obtained from the (VVV1, VVV2) vs. (2MASS, VVV2). The dispersion along the diagonal line is ($\sigma$$\approx$$8\deg$). The two small groups of points around the upper left and lower right corners are due because of the ``angle switch'' around 180$\degr$. }
\label{fig2}
\end{figure}

For the positional angle, the approximately small differences ($\sigma$$\approx$$8\deg$) produce the spread across the diagonal line. The two small groups of points around the upper left and lower right corners are due to the ``angle switch'' around 180$\degr$ (Fig.\ref{fig2}). There are very few outliers (i.e. most stars are spread along the
main diagonal). 

\begin{figure}
\includegraphics[width=\columnwidth]{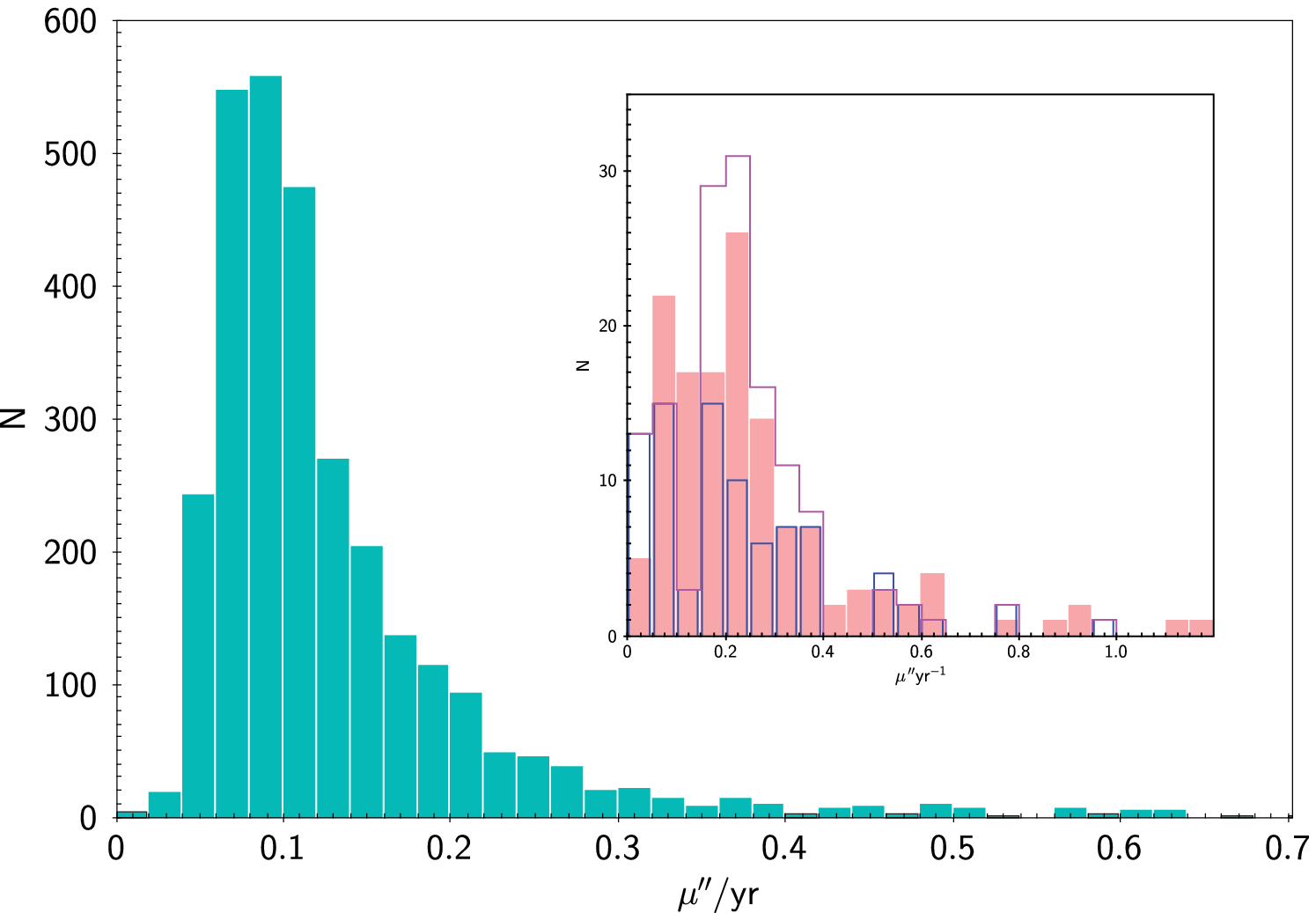}%
\caption{A histogram distribution of the modulus of the averaged (2MASS and VVV) PMs of the whole sample (green). The sample becomes significantly incomplete around PM$\sim$100 mas\,yr$^{-1}$. The inset gives the comparison of our PM sample brighter than $K_{\rm S}\leq9$ (red-filled histogram) with other two PM catalogues: \citet{2011AJ....142..138L} (blue open histogram) and the joined list of  \citet{2011AJ....142...10B} and \citet{2011AJ....142...92B} (purple open histogram). For details see Section~3.1. }
\label{fig3}
\end{figure}

The lower limit of the PM for our sample is about 20-30 mas\,yr$^{-1}$. The incompleteness of our sample becomes significant at  PM$\leq$100 mas\,yr$^{-1}$ (Fig.\ref{fig3}). One hundred thirty-four of the catalogue stars have PM$>$300 mas\,yr$^{-1}$ and 42 of them are newly found HPM objects. 382 catalogue stars have PM$>$200 mas\,yr$^{-1}$ and 179 of them are new HPM objects, 1576 stars showing PM>100 mas\,yr$^{-1}$ 1247 are new ones.  The last column of Table\ref{tab1}, labeled ``Comment'', gives short notes about known PM stars, binaries, and flags indicating the replacement of 2MASS with the VVV magnitudes and other useful information. 

\begin{figure*}
\includegraphics[width=\textwidth]{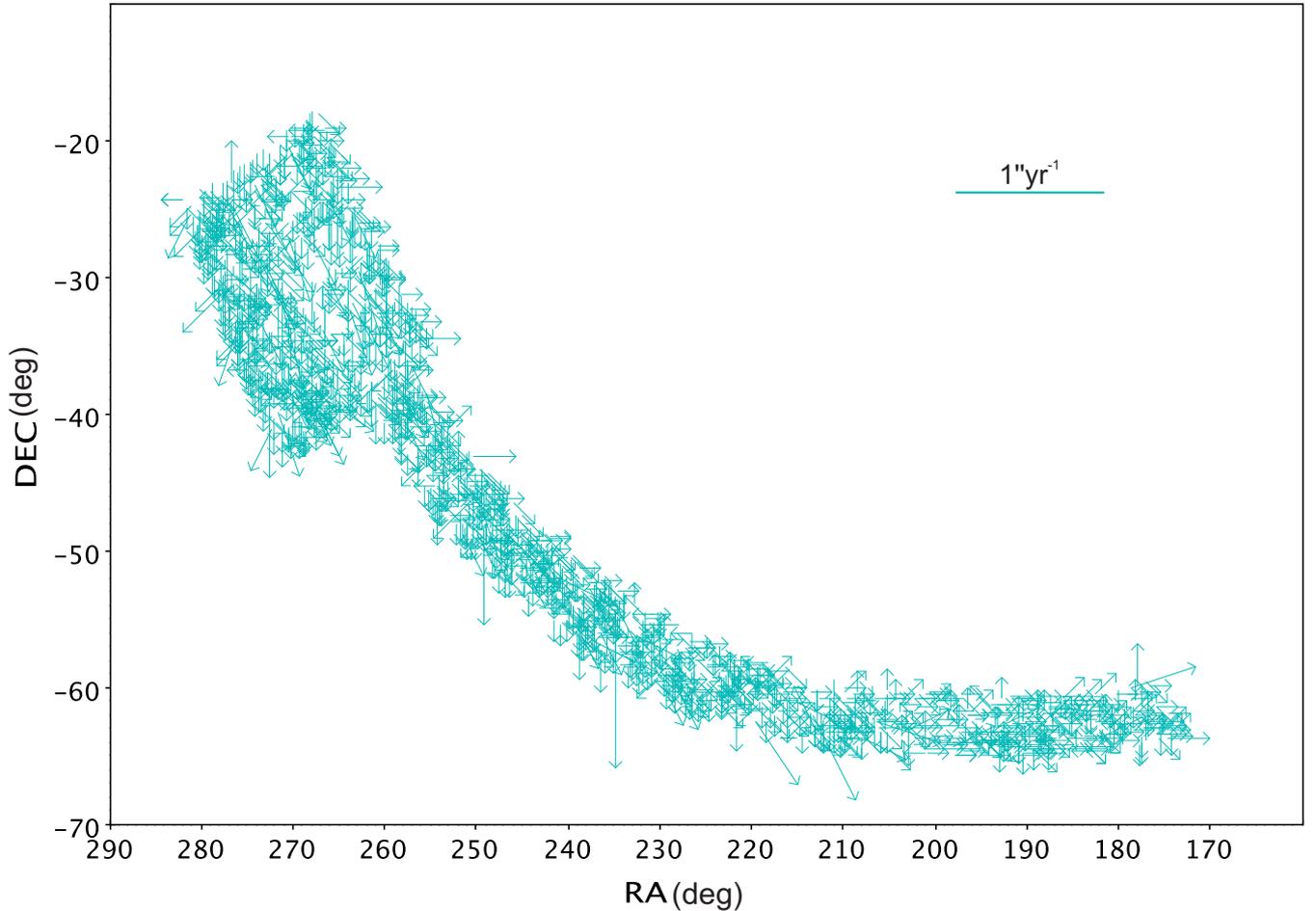}%
\caption{The distributions of the PM vectors over the VVV area. The modulus of a vector corresponding to a PM of 1 arsec\,yr$^{-1}$ is shown.}
\label{fig4}
\end{figure*}

\subsection{Catalogue caveats and completeness}
\label{sec:completeness}

As previously mentioned, bright sources are often either saturated or very close to saturation and their centroids fall in different places at different epochs and hence produce false HPMs for both non-PM and real PM stars.  

There is no single saturation magnitude for the VVV data, this varies slightly depending on different factors. The saturation limit depends on the seeing of the observations, as well as the DIT, and the filter used in the OBs. We also note that it is not the same for the different VIRCAM detectors that have different responses.  We take as $K_{\rm S}$=12.0 approximately as a representative value for the onset of the saturation in the VVV data. Saturated stars have a clear black dot centre in the images. We also find diffraction spikes of very bright stars as false HPM objects. Because of these reasons, we missed some known very bright HPM stars which were manually added to our list. Also, for some of the densest regions near the galactic centre we missed some HPM objects. These zones show clearly a lack of objects (see the vector diagram in Fig.\ref{fig4}).

It is tricky to estimate the completeness of a PM catalogue in the densest stellar regions. In order to do this we compared directly our catalogue with other PM catalogues covering the VVV area. The catalogues of \citet{2011AJ....142...10B}, \citet{2011AJ....142...92B} and \citet{2011AJ....142..138L} are shallower than our ($J,K_{\rm S}<9$) but are suitable to evaluate the completeness at the brightest end of our catalogue. The histogram is given in the inset of Fig.\ref{fig3}. The population of the comparison surveys is cut to the part falling within the VVV area. Our PM sample is cut to the limiting magnitude of ($K_{\rm S}\leq9$). The immediate analysis of the histogram shows that the joined list of  \citet{2011AJ....142...10B} and \citet{2011AJ....142...92B} catalogues (purple open histogram) is a bit well populated around the proper motion $\mu\sim0.2-0.4$ mas\,yr$^{-1}$ but the similarity between the two histograms leads us to conclude that the completeness of our catalogue in its brighter part is very similar to the completeness of the Boyd's catalogues. The \citet{2011AJ....142..138L} catalogue (blue open histogram) has lower completeness in these most crowded regions of the Milky way. The numbers are presented in the Table\,\ref{tab3}. The total number of 249 stars is a simple sum of non-repeated stars in the three catalogues. It means that the given completeness should be taken only as a upper limit.

\begin{table}
\caption{Comparison with \citet{2011AJ....142...10B}, \citet{2011AJ....142...92B} and \citet{2011AJ....142..138L} within VVV area. The total number of 249 stars is a simple sum of non-repeated stars of the three catalogues. }\label{tab3}
\centering%\small
\tabcolsep=1.5pt
\begin{tabular}{r|r|r}
\hline
  & & \\[-7pt]
  \multicolumn{1}{c|}{Survey/Catalog} & \multicolumn{1}{c|}{No stars} & \multicolumn{1}{c}{Completeness (\%)} \\
    &&\\[-7pt]
\hline

\citet{2011AJ....142...10B} and \citet{2011AJ....142...92B} & 136 & 55 \\
\citet{2011AJ....142..138L} & 86 & 35 \\
VVV PM & 136 & 55\\
Total & 249 & 100\\

\hline
\end{tabular}
\end{table}

\subsection{Reduced proper motion. Giants vs. dwarfs}
\label{sec:RPM}

\begin{figure}
\includegraphics[width=\columnwidth]{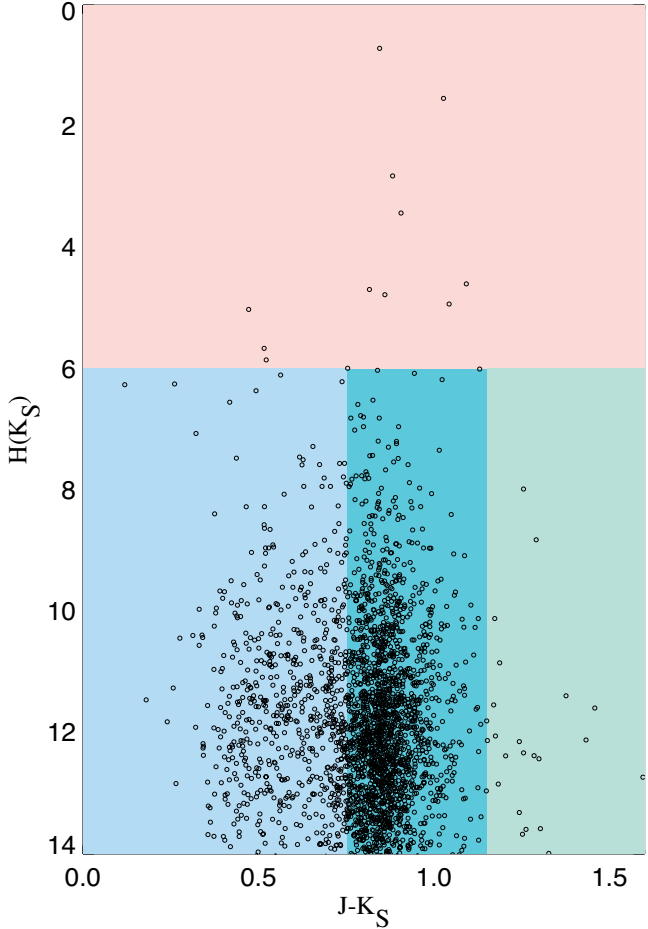}%
\caption{RPM versus the $J-K_{\rm S}$ colour index for the PM stars. The border between the reddish region and the three bluish-green regions at ${\rm H}(K_{\rm S})=6.0$ is the RPM cut used to separate the giants from the dwarfs. The left bluish region is populated by the earlier (K and few G dwarfs). The middle dark blue-green region contains the M-dwarfs and the right green region contains the later type (probably early L) objects.}
\label{fig5}
\end{figure}

\begin{figure}
\includegraphics[width=\columnwidth]{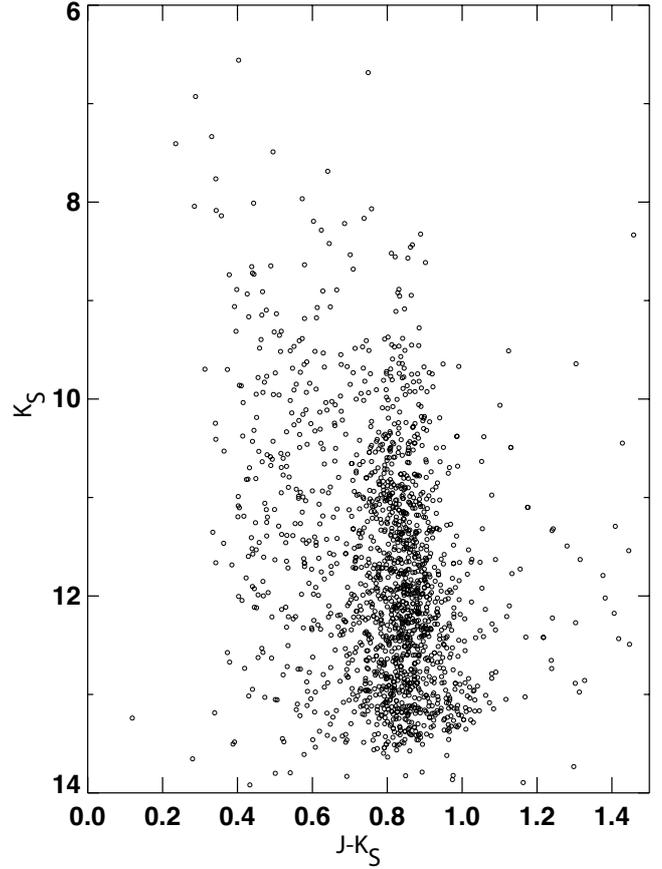}%
\caption{The $K_{\rm S}$ vs. $J-K_{\rm S}$ colour-magnitude diagram of all catalogue stars. }\label{fig6}
\end{figure}

When the distance to an object is not known, we can use the RPM, ${\rm H}$ in $K_{\rm S}$ filter, for the purpose of separating dwarfs from giant stars, where ${\rm H}(K_{\rm S})$ is used as a proxy for absolute magnitude. 
$${\rm H}(K_{\rm S})=K_{\rm S}+5\log\mu+5 $$ 
where $K_{\rm S}$ is the observed magnitude and $\mu$ is the observed PM in arcsec\,yr$^{-1}$. Since the reduced PM is analogous to absolute magnitude, a plot of H along with a colour index is a pseudo-equivalent of an Hertzprung-Russel diagram, and the giants are separated vertically from the dwarfs. \citet{2013MNRAS.435.2161F} found a clear separation between the dwarf and the giant populations at ${\rm H}_{K_{\rm S}}=6.0$. The giants populated the region above and the dwarfs below this limit on the RPM diagram. In our catalogue we have only a dozen objects, which lie in the region of the giants (red area in the Fig.\ref{fig5}) and a dozen more near the border between the two regions. 
These can be genuine giants with measured PMs or just relatively low PM dwarfs scattered there. For comparison a plot of the $K_{\rm S}$ vs. $J-K_{\rm S}$ colour-magnitude diagram is presented on Fig.\ref{fig6}. Finally, it is an expected result to have very few giants in our sample because the HPM objects in general should be nearby and if they are giants they should be very bright. 

\begin{figure}
\includegraphics[width=\columnwidth]{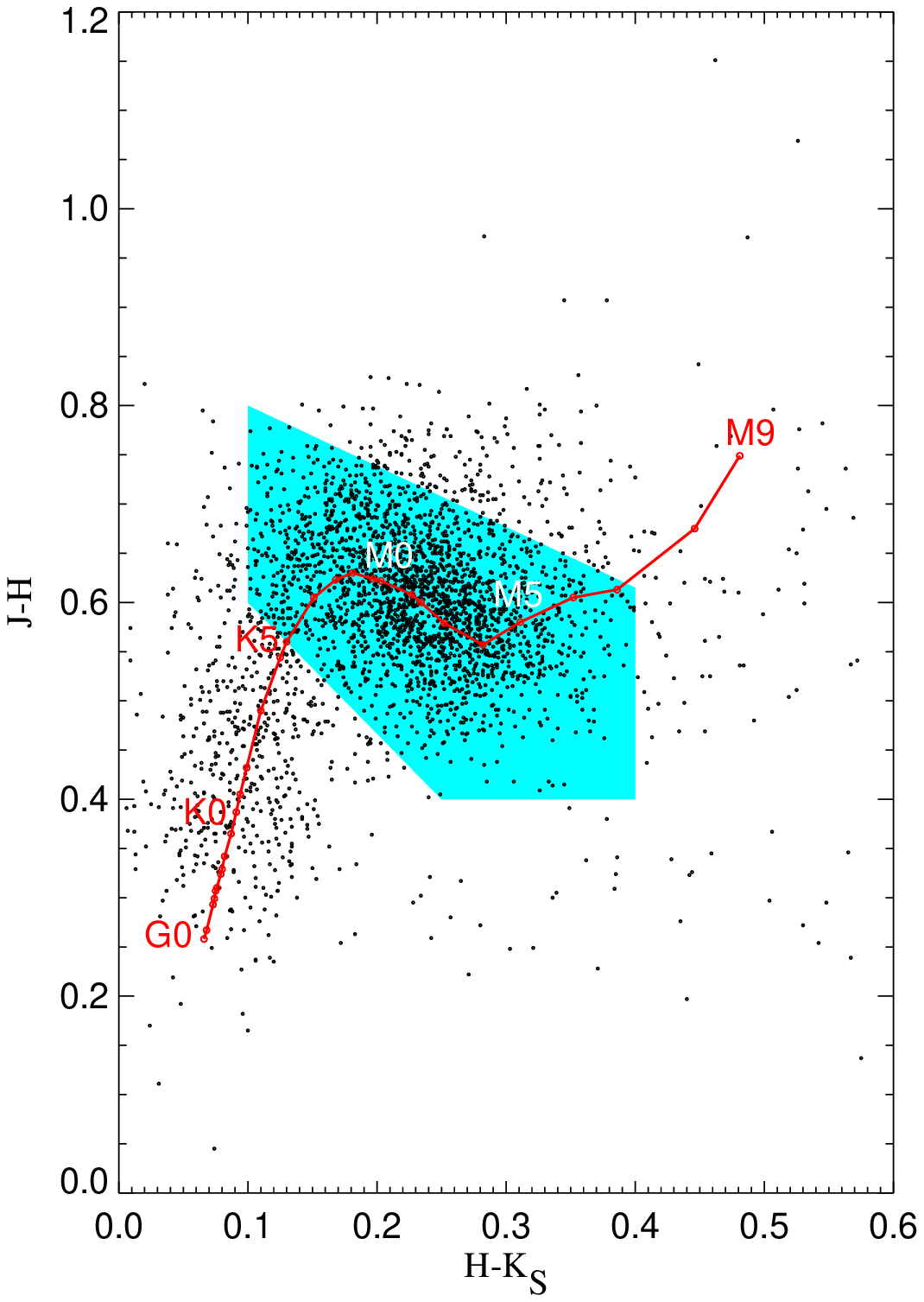}%
\caption{The $J-H$ vs. $H-K_{\rm S}$ colour-colour diagram of all catalogue stars. The Main sequence from G0 to M9 is given in red. The blue shadowed pentagon encloses the region of the M-dwarfs according to \citet{2011AJ....142..138L}.}
\label{fig7}
\end{figure}

\subsection{Nearby M-dwarfs}
\label{sec:m-dwarfs}
The majority of the PM objects in our catalogue are early- to mid-type M-dwarfs as can be seen in Fig.\ref{fig6} and Fig.\ref{fig7}. In fact M-dwarfs appear into a relatively compact region around $J-$$H$=0.6 and $H-$$K$=0.25. A small number of M-dwarfs fall outside this cluster of points; some have red $J-$$H$ colours more consistent with the M giants, but this most probably is an effect of photometric errors (see again the RPM diagram in Fig.\ref{fig5}). There are also few later spectral type objects, as well as numerous new early K- and few G-dwarfs with bluer colours. We are currently investigating the nature of the few remaining objects that fall outside the colours expected for normal main sequence dwarfs.

\section{Common proper motion binaries (CMPBs)}
\label{sec:cpmb}

Wide (100 AU) binary companions have long been used as a tool for identifying and studying faint stellar and substellar objects. Such systems are relatively common and $\sim$25\% of the solar type stars have companions wider than 100 AU \citep{2010ApJS..190....1R}. These objects are an important population for understanding models of binary star formation. Wide binaries also provide test cases for characterising stellar and substellar properties. As these systems likely formed from the same birth place (cluster or association), the companions should have the same metallicity and age as their host stars. For example, wide M dwarf companions to FGK stars have been used as calibrators for spectroscopic determinations of M dwarf metallicity relations. This kind of ``benchmark'' objects are even more important for characterising the substellar regime as the brown dwarfs lack a stable internal energy source and hence exhibit a degeneracy between their mass, luminosity and age. For substellar companions this degeneracy can be broken using the age and metallicity of the primary and the rest of the astrophysical parameters could be obtained from the spectra of the secondary and evolutionary models.

Identification of a common proper motion (CPM) and a common distance is necessary to link multiple stars as single, gravitationally bound systems. In an attempt to recover CPM companions we searched our catalogue for nearby objects within a radius of 3 arcmin, with PM difference less than 20 mas\,yr$^{-1}$, and a difference of the of the proper motion position angle within 10$^\circ$. There are 46 pairs which fulfil these criteria. We added to the list three additional pairs which do not fulfil the PM criterium and two additional ones which do not fulfil the positional angle criterium, but they are so close on the sky (< 8 arcsec) that there is no doubt they form physically connected systems. \cite{2013AA...560A..21I} studied a few of these common high proper motion binaries in the VVV data; the saturation and in some cases close proximity to the stars, adds to the proper motion errors and may
result in true common proper motion pairs having significantly different measured proper motions. During the visual inspection of all PM candidates we had overlooked 6 additional CPM systems with a secondary component fainter than the limiting magnitude of our search of $K_{\rm S}=13.5$ or very close to the primary and without photometry in CASU catalogs. For those close systems we provided additional PSF photometry of the secondary using the VISTA $JHK_{\rm S}$ images. The photometry was carried out with {\sc allstar} in {\sc daophot ii} \citep{1987PASP...99..191S}. The photometric calibration to the standard VISTA photometric system was performed by comparing our psf-magnitudes with CASU magnitudes of the stars around the targets. This is noted as {\sl not in the VVV catalogs, additional psf photom.} in the column ``Comment''.

At the end we generated a catalogue of 57 CPMBs given in  Table\ref{tab4}. The geometric criterion (of CPM) is clear, but to estimate the distances of individual components is difficult. For our CMPB candidates we have no parallax distances and we tried to obtain their photometric distances. For the brightest stars we searched the Optical catalogues (e.g. NOMAD, PPMX, USNO-A,-B, UCAC-2,-3, among others) and found their optical counterparts. Then applied the new set of photometric colour-$M_{K_{\rm S}}$ relations given in \citet{2014AJ....148..119F}. The uncertainty is due in some cases to false cross-identification between IR and optical sources and sometimes due to the crowding and the big pixel problem in the optical wavelengths. Then, we cross-identified the CPMBs with the VPHAS+ $ugri{\rm H}_\alpha$ optical catalogue \citep{2014MNRAS.440.2036D}. We found data for 37 of 107 individual stars in our CPMB catalogue (see Table\ref{tab4}). For these stars we also applied some of the photometric relations from \citet{2014AJ....148..119F}. For the non-saturated stars in the IR we also applied the \citet{2014A&A...571A..36R} colour-based spectral subtype and absolute magnitude calibration for M-dwarfs in the $YJHK_{\rm S}$ VISTA system. The final distance to the system is the average with MINMAX rejection of all estimations from the optical and IR calibrators and for both components. Once knowing the distance we converted the projected separation of the systems on the sky from arcsec to A.U., and using the apparent $K_{\rm S}$ magnitude we obtained the absolute magnitude $M_{K_{\rm S}}$ and then the approximate spectral type (with precision $\pm$ 1 subtype) of each star using the relations given in \citet{2013ApJS..208....9P}\footnote{(http://www.pas.rochester.edu/$\sim$emamajek/EEM\_dwarf\_UBVIJHK \_colors\_Teff.txt)}.  

The distances to these 57 systems vary from 20 to 310~pc with two potentially very nearby common proper motion pairs with distances $<$30 parsecs. The majority of the systems are M+M binaries. There are also four potential dK+dM and four dK+dK binaries. One of the systems, VVV~J141420.912-602336.19, has a faint companion which is a white dwarf, most probably of ZZ\,Cet type \citep{gromadzki2015}. In our PM search we found also a CPM brown dwarf companion to the A3V star $\beta$\,Circini. A complete follow-up and analysis of this very interesting ``benchmark'' object can be found in \citet{2015MNRAS.454.4476S}.

Table\ref{tab4} gives the mean parameters of the CPMB systems. The full table with the optical VPHAS+ and IR VVV magnitudes is available in electronic format only.

\begin{enumerate}
%\item Consecutive number
\item Right ascension for epoch 2000 
\item Declination for epoch 2000
\item The proper motion $\mu_\alpha\cos\delta$ in arsec\,yr$^{-1}$
\item The proper motion $\mu_\delta$ in arsec\,yr$^{-1}$
\item The total proper motion $\mu$ in arsec\,yr$^{-1}$
\item Positional angle of the proper motion PA in degrees
\item The projected separation of the system in arcsec
\item The projected separation of the system in A.U.
\item Distance in pc.
\item $K_{\rm S}$ magnitude
\item $(J-K_{\rm S})$ colour
\item Absolute magnitude $M_{K_{\rm S}}$
\item The approximate spectral type
\item Comment 
\end{enumerate}

\begin{landscape}
\begin{table}\tabcolsep=5pt
\caption{Common proper motion binary candidates. The full table with the optical VPHAS+ magnitudes is available in electronic format.}
\label{tab4}
\begin{tabular}{rrr|rrr|r|r|r|r|r|r|r|r|r|r|l|l}
\hline
&&&&&&&&&&&&&&&&&\\[-7pt]
  \multicolumn{3}{c|}{$\alpha(2000)$} &
  \multicolumn{3}{|c|}{$\delta(2000)$} &
  \multicolumn{1}{c|}{$\mu_\alpha\cos\delta$} &
  \multicolumn{1}{c|}{$\mu_\delta$} &
  \multicolumn{1}{c|}{$\mu$} &
  \multicolumn{1}{c|}{PA} &
  \multicolumn{1}{c|}{$\rho$} &
  \multicolumn{1}{c|}{$\rho$} &
  \multicolumn{1}{c|}{$r$} &
  \multicolumn{1}{c|}{$K_{\rm S}$} &
  \multicolumn{1}{c|}{$J$$-K_{\rm S}$} &
  \multicolumn{1}{c|}{$M_{K_{\rm S}}$} &
  \multicolumn{1}{c|}{SP} &
  \multicolumn{1}{c}{Comment}\\
  \multicolumn{1}{c}{(h)} &
  \multicolumn{1}{c}{(m)} &
  \multicolumn{1}{c}{(s)} &
  \multicolumn{1}{|c}{$(^\circ$)} &
  \multicolumn{1}{c}{$(')$} &
  \multicolumn{1}{c|}{$('')$} &
  \multicolumn{1}{c|}{$(''{\rm yr}^{-1}$)} &
  \multicolumn{1}{c|}{$(''{\rm yr}^{-1}$)} &
  \multicolumn{1}{c|}{$(''{\rm yr}^{-1}$)} &
  \multicolumn{1}{c|}{($^\circ$)} &
  \multicolumn{1}{c|}{$('')$} &
  \multicolumn{1}{c|}{(AU)} &
  \multicolumn{1}{c|}{(pc)} &
  \multicolumn{1}{c|}{(mag)} &
  \multicolumn{1}{c|}{(mag)} &
  \multicolumn{1}{c|}{(mag)} &
  \multicolumn{1}{c|}{} &
  \multicolumn{1}{c}{}  \\
  \hline
  &&&&&&&&&&&&&&&&&\\[-7pt]
  11 & 36 & 59.321 & -63 & 29 & 41.89 & -0.127 & -0.047 & 0.135 & -110.3 & 45.15 & 3341 & 74 & 11.906 & 0.919 & 7.56 & M4 & COMP A\\
  11 & 36 & 59.864 & -63 & 28 & 56.89 & -0.115 & -0.048 & 0.125 & -112.7 &  &  & 74 & 12.358 & 0.89 & 8.012 & M5 & COMP B\\
  &&&&&&&&&&&&&&&&&\\[-7pt]
  11 & 39 & 48.022 & -62 & 52 & 33.44 & -0.16 & 0.028 & 0.162 & -80.1 & 1.25 & 180 & 144 & 9.717 & 0.708 & 3.918 & K1 & COMP A\\
  11 & 39 & 47.951 & -62 & 52 & 32.01 & -0.163 & 0.021 & 0.162 & -82.3 &  &  & 144 & 12.535 & 0.599 & 6.736 & M3 & COMP B, not in the VVV catalogs, additional psf photom. \\
  &&&&&&&&&&&&&&&&&\\[-7pt]
  11 & 48 & 25.132 & -62 & 42 & 54.22 & -0.063 & 0.032 & 0.071 & -63.1 & 1.8 & 259 & 144 & 13.394 & 0.725 & 7.598 & M4 & COMP A \\
  11 & 48 & 25.028 & -62 & 42 & 55.52 & -0.069 & 0.030 & 0.073 & -60.2 &  & & 144 & 15.259 & 0.745 & 9.463 & M6 & COMP B, not in the VVV catalogs, additional psf photom. \\
  &&&&&&&&&&&&&&&&&\\[-7pt]
  12 & 29 & 29.542 & -62 & 9 & 39.67 & -0.123 & 0.046 & 0.131 & -69.5 & 13.13 & 696 & 53 & 10.177 & 0.899 & 6.556 & M3 & COMP A \\
  12 & 29 & 31.117 & -62 & 9 & 32.55 & -0.123 & 0.047 & 0.132 & -69.1 &  &  & 53 & 13.175 & 0.861 & 9.554 & M6 & COMP B\\
&&&&&&&&&&&&&&&&&\\[-7pt]   
  \multicolumn{18}{l}{... And 53 more CPM binaries ...} \\
  &&&&&&&&&&&&&&&&& \\[-7pt]
\hline
\end{tabular}
\end{table}
\end{landscape}

\section{Summary}

This work has identified 3003 PM stars (mostly late K- and M-dwarfs) with magnitudes of $K_{\rm S}$<13.5 and PM$<$30~mas\,yr$^{-1}$ from the VVV catalogues. The completeness in its brighter part ($K_{\rm S}$<9) is comparable with the completeness of the similar catalogues outside the ``zone of avoidance'' near to the MW bulge and disk. We found 57 wide CPMBs all dK+dM or dM+dM binaries. We started an intense spectral follow-up of the most interesting candidates \citep{gromadzki2015}. Low-resolution spectra will confirm the spectral types of the objects and higher-resolution spectra will provide constraints on the RV. Such observations would allow prioritisation of bright younger M-dwarfs for light-curve follow-up and transit searches and also for AO imaging for searching for nearby companions. 

The incompleteness of our sample becomes significant at  PM$\leq$100 mas\,yr$^{-1}$. One hundred thirty-four of the catalogue stars have PM$>$300 mas\,yr$^{-1}$ and 42 of them are newly found HPM objects. 382 catalogue stars have PM$>$200 mas\,yr$^{-1}$ and 179 of them are new HPM objects, 1576 stars showing PM>100 mas\,yr$^{-1}$ -- 1247 are new ones. The star with the highest proper motion in the VVV area (except $\alpha$~Centauri) is HD~156384C with PM=1175 mas\,yr$^{-1}$ and the new found star with the highest PM motion is VVV~J180414.62-312937.18 with PM=810 mas\,yr$^{-1}$. 

Here we limited our search only for the brightest objects. The three thousand HPM stars found are only the tip of the iceberg. Most tiles in the VVV database have limiting magnitudes $K_{s,{\rm lim}}\sim17$--18 \citep[see the VVV DR1 paper][]{2012A&A...544A.147S}, and we expect that the final catalogue of VVV HPM stars will contain $>$10$^5$ objects.

\section*{Acknowledgements} 
This project is supported by the Ministry of Economy, Development, and Tourism's Millennium
Science Initiative through grant IC120009, awarded to The Millennium Institute of Astrophysics, MAS. Support for RK is provided from Fondecyt Reg. No. 1130140. RK,  MG, JB, PL, LS, MK and AY acknowledge support from CONICYT REDES project No. 140042. MG acknowledges support also from Joined Committee ESO and Government of Chile 2014. J.C.B. acknowledges support from CONICYT FONDO GEMINI - Programa de Astronom\'ia del DRI, Folio 32130012. AA acknowledges support from SU-NSF 81/2016 grant.

This publication makes use of data products from the Two Micron All Sky Survey, which is a joint project of the University of Massachusetts and the Infrared Processing and Analysis Center/California Institute of Technology, funded by the National Aeronautics and Space Administration and the NationalScience Foundation. 

This research has made use of the SIMBAD database, operated at CDS, Strasbourg, France

\end{document}